\documentclass[aps,prb,twocolumn,showpacs,showkeys,footinbib,amsmath,amssymb,superscriptaddress,longbibliography]{revtex4-2}
\usepackage{hyperref}
\usepackage{amsfonts}
\usepackage{graphicx}
\usepackage{bm}
\usepackage{relsize}
\usepackage{braket}
\usepackage{bbold}
\usepackage{float}

\begin{document}

\title{Emergent bright excitons with Rashba spin-orbit coupling in atomic monolayers}

\author{Jiayu David Cao}
\email{jiayucao@buffalo.edu}
\affiliation{Department of Physics, University at Buffalo, State University of New York, Buffalo, New York 14260, USA}

\author{Gaofeng Xu}
\affiliation{Department of Physics, Hangzhou Dianzi University, Hangzhou, Zhejiang 310018, China}

\author{Benedikt Scharf}
\affiliation{ Institute for Theoretical Physics and Astrophysics and W{\"u}rzburg-Dresden Cluster of Excellence ct.qmats,
University of W{\"u}rzburg, Am Hubland, 97074 W{\"u}rzburg, Germany}

\author{Konstantin Denisov}
\affiliation{Department of Physics, University at Buffalo, State University of New York, Buffalo, New York 14260, USA}

\author{Igor \v{Z}uti\'c}
\email{zigor@buffalo.edu}
\affiliation{Department of Physics, University at Buffalo, State University of New York, Buffalo, New York 14260, USA}
\date{\today}
\begin{abstract}
Optical properties in van der Waals heterostructures based on monolayer transition-metal dichalcogenides (TMDs), are often dominated by excitonic transitions. While intrinsic spin-orbit coupling (SOC) and an isotropic band structure are typically studied in TMDs, in their heterostructures Rashba SOC and trigonal warping (TW), resulting in bands with threefold anisotropy, are also present. By considering a low-energy effective Hamiltonian and Bethe-Salpeter equation, we study the effect of Rashba SOC and TW on the band structure and absorption spectra. Rashba SOC is predicted to lead to emergent excitons, which are identified as an admixture between $1s$ and $2p$ symmetries. In contrast, for experimentally relevant values, TW has only a negligible effect on the absorption spectrum. These findings could guide experimental demonstrations of emergent bright excitons and further studies of the proximity effects in van der Waals heterostructure. 
\end{abstract}
\maketitle

\section{Introduction}

There is a growing class of atomically-thin van der Waals (vdW) materials with striking optical properties which, unlike their bulk counterparts,
can be highly tunable and have strong Coulomb 
interactions~\cite{Wang2018:RMP,Carvalho2016:NRM,Tang2022:PSSB,Syperek2022:NC,Wozniak2023:S,Rudenko2022:PRB,Grzeszczyk2023:AM,Qiu2013:PRL}. 
These properties are exemplified in transition-metal dichalcogenides (TMDs) with
direct band gaps and efficient light emission, valley-dependent helicity of optical transitions, broken inversion symmetry together with a strong intrinsic spin-orbit coupling (SOC),
and tightly-bound excitons, dominating the optical response~\cite{Xiao2012:PRL,Chernikov2014:PRL,Amani2015:S,Mak2016:NP}.

In vdW heterostructures, as a 
platform to transform their atomically-thin materials through proximity effects, the role of TMDs is twofold. TMDs acquire magnetic, superconducting, enhanced SOC, charge-density wave, or topological 
properties~\cite{Zhao2017:NN,Zhong2017:SA,Zutic2019:MT,Joshi2022:APLM,Bora2021:JPM,Zhong2020:NN,Zhang2022:PRB,Saito2016:NP,Cortes2019:PRL,Zatko2019:ACSN}, 
while they also transform the neighboring regions, for example, by altering their magnetic properties~\cite{Tu2022:APL,Dolui2020:PRM}. Since the resulting change in the excitonic spectrum of atomically-thin materials is a probe for magnetic proximity effects or a tunable band topology~\cite{Zhang2018:PRL,Scharf2017:PRL,Xu2020:PRL,Gloppe2022:PRB,Henriques2020:PRB,Onga2020:NL,Dias2020:PRB,Wang2022:npjCM}, it is crucial to establish 
a description of these systems beyond the single-particle picture~\cite{Scharf2017:PRL}.

\begin{figure}[h]
	\vspace{-0.4cm}
	\centering
	\includegraphics[width=0.46
	\textwidth]{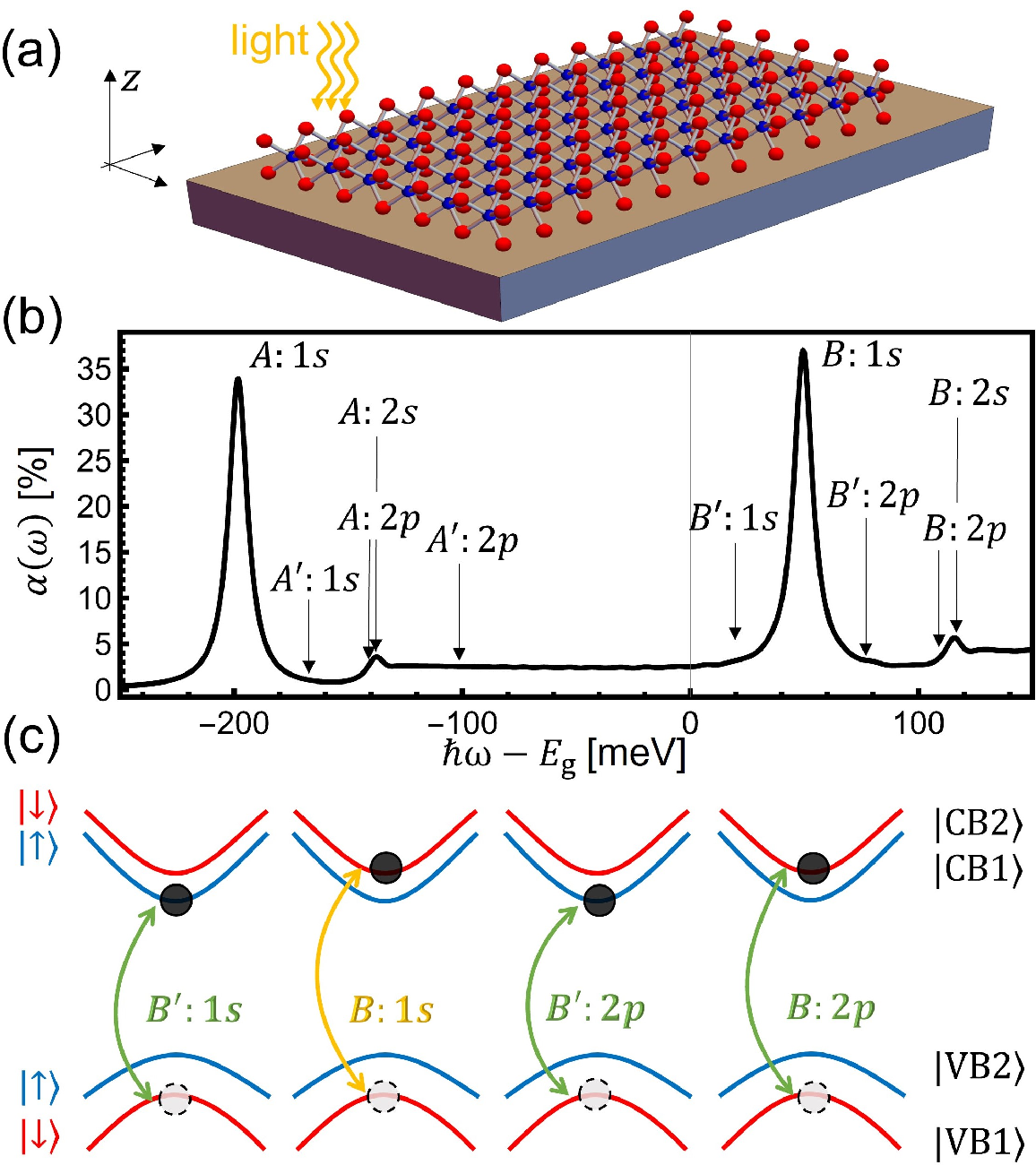}
	\caption{\label{fig:FR1} (a) Schematic of an atomic ML on a substrate. (b) Absorption spectrum computed numerically without trigonal warping or Rashba SOC, based on the Bethe-Salpeter equation described in Sec.~\ref{sec:II theoretical Model}.  
				The absorption peaks denote various excitons:  $A$ ($A'$) are formed from CB1 (CB2) and VB2, while $B$ ($B'$) are formed from CB2 (CB1) and VB1 
		where CB1,2 (VB1,2) are conduction (valence) subbands with spins marked by arrows. 
		(c) Examples of $B$ and $B'$ excitons in ascending energy order in the $K$ valley. Yellow (green) arrows: bright (dark) excitons. Black (gray) circles: electrons (holes). }
		\vspace{-0.5cm}
\end{figure}

In this work, we focus on the evolution of the optical absorption in monolayer (ML) TMDs 
with Rashba SOC and trigonal warping (TW), which are inherent to TMDs and their vdWs heterostructures, but whose combined role on excitons remains largely unexplored~\cite{Wang2018:RMP}. Similar to the extensively studied two-dimensional electron 
gas~\cite{Bercioux2015:RPP,Zutic2004:RMP}, Rashba SOC  occurs in TMD systems with broken structural inversion symmetry and its strength can be 
tuned by an externally applied electric field. In TMDs this
symmetry breaking is readily realized due to a substrate [shown in Fig.~\ref{fig:FR1}(a)]
or in ML Janus TMDs~\cite{Tang2022:PSSB,Chen2020:RSCA}. 

The presence of TW in TMDs removes the usually assumed isotropic dispersion near the $K$/$K'$ valleys 
and introduces a threefold
symmetry accompanied with strong nonlinear optical properties~\cite{Mahmoudi2022:PRB,Saynatjoki2017:NC,Zhumagulov2022:PRB}. 
We base our approach on
the Bethe-Salpeter equations, to include 
strong Coulomb interaction that, with reduced screening, 
can lead to
exciton binding energies of up to 0.5~eV, orders of magnitude larger than in bulk semiconductors~\cite{Wang2018:RMP}.

In Fig.~\ref{fig:FR1}(a) we illustrate the considered system,
illuminated by light. The corresponding absorption spectrum in Fig.~\ref{fig:FR1}(b), shown without Rashba SOC and TW,
has two main peaks indicating the $A$ and $B$ excitons, bound electron-hole pairs that can be described within the low-energy
four-band model, shown in Fig.~\ref{fig:FR1}(c) for a single valley. In  semiconducting ML TMDs (MX$_2$, M = Mo, W; X = S, Se, Te), the SOC
splitting is larger for the valence band (VB) than the conduction band (CB), of which the spin ordering is opposite for M= Mo and W~\cite{Xiao2012:PRL,Wang2018:RMP}. 

The lower energy A exciton corresponds to the bound state formed from the hole in the upper valence subband (VB2)
and the electron in the lower conducting subband (CB1), while the B exciton is formed from the VB1 and CB2 subbands, shown in the 2nd diagram in Fig.~\ref{fig:FR1}(c).
Unlike these bright A and B excitons that correspond to the optically-allowed dipole transitions in Fig.~\ref{fig:FR1}(c),   $2p$ 
excitons (3rd, 4th diagram) and spin-forbidden B' series excitons (1st , 3rd diagram) are optically forbidden and termed dark.
Here the notation for various excitons,  $1s$, $2p$,... is derived from their similarity with the hydrogen-like states below the band gap~\cite{Klingshirn:2012}, 
while the CB ordering corresponds to MoX$_2$ TMDs.

Our calculated results for the absorption spectrum in Fig.~\ref{fig:FR1} contain basic properties about the bright and
dark excitons in TMDs and provide a reference point to identify a possible influence of Rashba SOC and TW. 
While Rashba SOC is absent in nonpolar ML TMDs (MX$_2$, M = Mo, W; X = S, Se, Te), where the two planes of X chalcogen atoms are symmetrically
surrounding the plane of M metallic atoms, it is inherent and can be very strong in polar ML Janus TMDs (MXY, M = Mo, W; X = S, Se, Te; Y $\neq$ X),
induced by the broken out-of-plane mirror symmetry~\cite{Tang2022:PSSB}. TW is predicted for graphene and its 
multilayers~\cite{Ando1998:JPSJ,Dresselhaus2002:AP}, as well as in ML TMDs and Janus TMDs~\cite{Rostami2013:PRB,Kormanyos2013:PRB,Kormanyos2015:2DM,Korkmaz2021:JPCC}. However, in studies of excitons in atomic MLs with a hexagonal lattice, the band structure at the $K$/$K'$ valleys is usually modeled assuming an isotropic dispersion which neglects TW.

An additional motivation for our work comes from recognizing that excitonic features in atomic MLs provide 
fingerprints for proximitized materials, for example, modified by magnetism, charge density wave, or tunable band topology~\cite{Scharf2017:PRL,Joshi2022:APLM,Xu2020:PRL}. Our findings on the role of Rashba SOC and TW could also 
guide future studies in more complex systems including the interplay of multiple proximity effects. 
Even for seemingly well-understood magnetic proximity effects, the common picture is incomplete. 
In contrast to the assumed similarity of magnetic proximity with an applied magnetic field leading to a spin splitting in the TMD, 
the modification of the excitonic spectrum in MoSe$_2$ placed on top of a magnetic insulator, CrBr$_3$, 
reveals a valley-dependent shift indicating that the proximity-induced spin splitting is valley dependent~\cite{Choi2023:NM,Zhou2023:NM}. Further measurements
of excitonic features of other atomic MLs may reveal even more surprises. 

Following this Introduction, in Sec.~\ref{sec:II  theoretical Model} we outline our theoretical framework using a low-energy Hamiltonian and  Bethe-Salpeter equation to describe excitons in ML TMDs. In Sec.~\ref{sec:III results} we discuss the results for our computational system, based on ML MoTe$_2$ on a substrate with a large dielectric constant. Specifically, we focus 
on the modifications of the optical properties with Rashba SOC and TW. While Rashba SOC can lead to pronounced effects and unexplored emergent excitons, 
the role of TW has only a very modest effect on the absorption spectra.
In Sec.~\ref{sec:IV Conclusions} we provide our conclusions and an outlook. 

\section{Theoretical Model \label{sec:II theoretical Model}}
\subsection{Low-energy effective Hamiltonian \label{sec:II: H} }

In the absence of  broken time-reversal symmetry and net magnetization, for a description of a ML TMD on a substrate as shown in Fig.~\ref{fig:FR1}(a), 
it is sufficient to consider one of the valleys, $K$, while the behavior of the other valley, $K'$, is related through the time-reversal operation.
For bands with a two-dimensional (2D) representation, the SOC Hamiltonian can be written as $H_\mathrm{SO}=\bm{\Omega}(\bm{k})\cdot\bm{s}$ using the SOC 
field $\bm{\Omega}(\bm{k})$~\cite{Zutic2004:RMP,Fabian2007:APS,Zutic2019:MT}, where $\bm{k}$ is the wave vector and $\bm{s}$ is the vector of spin Pauli matrices. 
In ML TMDs, this leads to $\bm{\Omega}(\bm{k})=\lambda(\bm{k})\hat{\bm{z}}$, where $\lambda(\bm{k})$ is odd in $\bm{k}$ and $\hat{\bm{z}}$ is 
the unit vector normal to the ML plane. At the $K$ point, $\lambda(\bm{k})$ reduces to the values $\lambda_\mathrm{c(v)}$ in the CB (VB).
Within the single-particle picture, the total low-energy Hamiltonian is given by~\cite{Kormanyos2015:2DM}
\begin{equation}
H_\mathrm{tot}=H_0+H_\mathrm{R}+H_\mathrm{TW},
\label{eq:Htot}
\end{equation}
which is a sum of the  ``bare" ML-TMD, Rashba SOC, and TW terms~\cite{Kormanyos2013:PRB,Kormanyos2015:2DM}, respectively. Here the first term in the four-band model that includes an inherent SOC can be expressed as     
\begin{multline}
 H_0 = \hbar v_\mathrm{F}(k_x \sigma_x \tau_z +k_y \sigma_y) + (E_\mathrm{g}/2)\sigma_z   \\
 +\tau_z s_z\left[\lambda_c(1+\sigma_z)/2+\lambda_v(1-\sigma_z)/2\right],
\label{eq:H0}
\end{multline}
where $\sigma_i$ and $\tau_i$ denote Pauli matrices for the CB or VB, and $K/K'$ valley, respectively. $v_F$ is the Fermi velocity, $k_{x,y}$ are wave vectors 
measured from $K/K'$, $E_\mathrm{g}$ is the band gap, 2$\lambda_c$ and 2$\lambda_v$
are the SOC splitting in the CB and VB. With $\lambda_R$  
and $\kappa$  describing the strength of Rashba SOC and TW,
we have
\begin{equation}
	H_\mathrm{R}=\lambda_\mathrm{R}(s_y \sigma_x \tau_z-s_x \sigma_y),\\
\label{eq:HR}
\end{equation}
and
\begin{equation}
	H_\mathrm{TW}=\kappa\left[(k_x^2-k_y^2)\sigma_x-\tau_z 2 k_x k_y \sigma_y\right].
\label{eq:HTW}	
\end{equation}
Equations (\ref{eq:Htot})-(\ref{eq:HTW}), describe the single-particle band structure of our system, which serves as a starting point for our many-body calculations.

\subsection{Bethe-Salpeter equation and excitons \label{sec:II:BSE}}
From the resulting single-particle description $H^\tau_\mathrm{tot}\eta^\tau_{n\bm{k}}=\epsilon_n^\tau(\bm{k})\eta^\tau_{n\bm{k}}$, with the eigenenergies $\epsilon_n^\tau$ and the corresponding eigenstates $\eta^\tau_{n\bm{k}}$ as the basis, we take into account many-body effects through the Bethe-Salpeter equation (BSE)~\cite{Hanke1975:PRB,Hanke1980:PRB,Rohlfing2000:PRB,Scharf2016:PRB,Scharf2017:PRL} to study excitons in the presence of Rashba SOC and TW,
\begin{equation}
	\left[\Omega^{S,\tau} - \epsilon_c^\tau(\bm{k}) + \epsilon_v^\tau(\bm{k}) \right]
	\mathcal{A}_{vc\bm{k}}^{S\tau}	=\sum_{v'c'\bm{k}'}\mathcal{K}_{vc\bm{k},v'c'\bm{k}'}^\tau \mathcal{A}_{v'c'\bm{k}'}^{S\tau},
\label{eq:BSE}	
\end{equation}
where in a given valley $\tau$ the band index $n=c(v)$ denotes one of the two CB (VB) subbands. $\Omega^{S,\tau}$ is the energy of an exciton in the state $S$, given by the 
wave function $|\Psi_S^\tau \rangle$ =$\sum_{vc\bm{k}} \mathcal{A}_{vc\bm{k}}^{S\tau} \hat{c}^{\dagger}_{\tau c\bm{k}} \hat{c}_{\tau v\bm{k}}|GS\rangle$ with coefficients $\mathcal{A}_{vc\bm{k}}^{S\tau}$, the creation (annihilation) operator of an electron in a CB $c$ (VB $v$) $\hat{c}^\dagger_{\tau c \bm{k}} (\hat{c}_{\tau v \bm{k}})$ in a valley $\tau$, and the ground state $|GS\rangle$ with fully occupied VBs and empty CBs.
The kernel $\mathcal{K}_{vc\bm{k},v'c'\bm{k}'}^\tau$ includes the Coulomb interaction between electrons in the layer, determined from the dielectric environment, geometry, and form factors calculated from 
$\eta_{n\bm{k}}^\tau$~\cite{Keldysh1979:JETP,Cudazzo2011:PRB,Scharf2017:PRL,MartinsQuintela2022:PSSB}.
While $\mathcal{K}^{\tau}_{vc\bm{k},v'c'\bm{k}'}=\mathcal{K}^{\mathrm{d},\tau}_{vc\bm{k},v'c'\bm{k}'}+\mathcal{K}^{\mathrm{x},\tau}_{vc\bm{k},v'c'\bm{k}'}$
consists of the direct and exchange terms~\cite{Rohlfing2000:PRB,Scharf2017:PRL}, since we do not consider inter-valley transitions, 
the exchange term vanishes due to the orthogonality of the eigenspinors $\eta_{n\bm{k}}$ \cite{Scharf2017:PRL, Xu2020:PRL}. 
The interaction kernel only contains the direct term
of the form~\cite{Scharf2017:PRL,Wang2018:RMP} 
\begin{equation}
	\mathcal{K}^{\tau}_{vc\bm{k},v'c'\bm{k}'}=-\frac{V\left(\bm{k}-\bm{k}'\right)f^\tau _{cc'}\left(\bm{k},\bm{k}'\right)f^\tau_{v'v}\left(\bm{k}',\bm{k}\right)}{A},
\label{eq:KerneltEffectiveModel}
\end{equation}
where
$V(\bm{q}\equiv\bm{k}-\bm{k'})$ is the Fourier transform of the bare Coulomb potential determined from the dielectric environment, while 
the form factors $f^{\tau}_{nn'}\left(\bm{k},\bm{k}'\right)=\left[\eta^{\tau}_{n\bm{k}}\right]^\dagger\eta^{\tau}_{n'\bm{k}'}$ are calculated from the single-particle states,
and A is the unit area.
In the absence of Rashba SOC, spin is a good quantum number, so the form factors vanish for coupling between different CBs (VBs), and the CBs (VBs) decouple in a kernel $\mathcal{K}^{\tau}_{vc\bm{k},v'c'\bm{k}'}$.  
The bare Coulomb potential is determined only from the dielectric environment
and can be obtained from the Poisson equation.
Correspondingly, 
$V(\bm{q})$ between two electrons in the $xy$-plane ($z=z'=0$) for a ML TMD of thickness $d$ is given by~\cite{Zhang2014:PRB,Scharf2017:PRL,Scharf2019:PRB}
\begin{equation}
		V(\bm{q})\approx
	\frac{2\pi e^2}{\varepsilon q+r_0q^2},
\label{eq:Coulomb_potential_approx}
\end{equation}
where this expression follows from an expansion in powers of $qd$, which is very accurate for a thin layer, $qd\ll1$. Here
$\varepsilon=(\varepsilon_t+\varepsilon_b)/2$
is the average (background) dielectric constant of the top and bottom materials surrounding the ML TMD.
The ML polarizability is
\begin{equation}
r_0=\frac{\tilde \varepsilon d}{2}\left(1-\frac{\varepsilon_t^2+\varepsilon_b^2}{2\tilde{\varepsilon}^2}\right),
\label{eq:r0}
\end{equation}
where $\tilde{\varepsilon}$ is the dielectric constant of the single atomic layer. 
In the limit of $\varepsilon_{t/b}\ll\tilde{\varepsilon}$, $r_0=\tilde{\varepsilon}d/2$ \cite{Keldysh1979:JETP,Cudazzo2011:PRB}. The interaction given by Eq.~(\ref{eq:Coulomb_potential_approx}) has proven to be 
successful 
in capturing the excitonic properties of ML-TMDs \cite{Berkelbach2013:PRB,Berkelbach2015:PRB}.

While excitons in TMDs deviate from 2D hydrogen due to screening, they can still be labeled as $1s$, $2s$, $2p$, etc., based on their 
symmetry~\cite{Qiu2016:PRB,Ye2014:N,Rohlfing2000:PRB,Berkelbach2015:PRB}. To classify the excitons according to their symmetries, 
we study the real-space exciton wave function
\begin{equation}
	\Psi^{S, \tau}(\bm{r_e}-\bm{r_h})=\sum_{vc\bm{k}} {\mathcal A}_{vc\bm{k}}^{S\tau} \psi^\tau_{c\bm{k}}(\bm{r_e})\psi_{v\bm{k}}^{\tau *}(\bm{r_h}),
\label{eq:wave_total}
\end{equation}
where $\psi^\tau_{c\bm{k}}({\bm{r_e}}) \psi_{v\bm{k}}^{\tau *}(\bm{r_h})=\mathrm{e}^{i\bm{k}\cdot(\bm{r_e}-\bm{r_h})}\sum_{i,j}[\eta_{c\bm{k}}\eta_{v\bm{k}}^\dagger]_{ij}/A$, and $\bm{r}_e$ ($\bm{r}_h$) are the electron (hole) coordinates. The single-particle eigenstate 
wave function can be calculated as $\psi^\tau_{n\bm{k}}(\bm{r})=\mathrm{e}^{i\bm{k}\cdot\bm{r}}\sum_m[\eta^\tau_{n\bm{k}}]_m\langle\bm{r}|m\bm{k}\rangle/\sqrt{A}$, where $m$ is the subband index. Since we focus on the exciton symmetry, we assume a constant periodic part $\langle\bm{r}|n\bm{k}\rangle$ of the Bloch function for the calculation. This simplification removes the crystal site dependence of the exciton wave function, consistent with our low-energy Hamiltonian. Thus the wave function in our work is actually the exciton envelope wave function~\footnote{Which is smooth and different from Figs.~3(b) and 3(c) in Ref.~[\onlinecite{Qiu2013:PRL}].}.
As noted for Eq.~(\ref{eq:KerneltEffectiveModel}), without Rashba SOC, the CBs (VBs) in $\mathcal{K}^{\tau}_{vc\bm{k},v'c'\bm{k}'}$ decouple. Thus $A$ and $B$ exciton series are independent. 

Unlike in 2D hydrogen, a state with higher angular momentum can have a larger binding energy than a state with lower angular momentum. For example, the first few low-energy states are $1s$, $2p$,  and $2s$,  as shown in Fig.~\ref{fig:FR1}(b). This different ordering is already well known~\cite{Berkelbach2015:PRB,Ye2014:N}
and is expected since the spatial dependence of the screened Coulomb interaction in TMDs deviates from the hydrogenic form.
However, TMD excitons obey identical selection rules as 2D hydrogen. 
For example, $s$-type excitons are bright, $p$-type excitons are dark~ \cite{Berkelbach2015:PRB}. Signatures of bright excitons are given in the absorption spectrum
\begin{equation}
	\alpha^\pm(\omega)=\frac{4 e^2 \pi^2}{ c\omega}\frac{1}{A}\sum_{S\tau} \left|\sum_{vc\bm{k}}\mathcal{D}_{vc\bm{k}}^{\tau,\pm}\mathcal{A}_{vc\bm{k}}^{S\tau}\right|^2\delta(\hbar\omega-\Omega^S),
\label{eq:abs}
\end{equation}
where $\omega$ is the photon frequency of light propagating along the $-\hat{z}$ direction, and $c$ is the speed of the light.
The single-particle velocity matrix elements are given by $\mathcal{D}^{\tau, \pm}_{vc\bm{k}}=\left[\eta^{\tau}_{v\bm{k}}\right]^\dagger\hat{v}_\pm\eta^{\tau}_{c\bm{k}}$ for left/right circularly polarized light  (helicity eigenstates) 
with $\hat{v}_\pm=(\hat{v}_x\pm i \hat{v}_y)/\sqrt{2}$, $\hat{v}_{x/y}=\partial H_\mathrm{tot}/\partial(\hbar k_{x/y})$. A zero sum $\sum_{vc\bm{k}} \mathcal{D}_{vc\bm{k}} \mathcal{A}^{S\tau}_{vc\bm{k}}$ indicates a dark exciton. The $\delta$ function is modeled by a Lorentzian with broadening $\Gamma$, where a smaller $\Gamma$ is necessary to 
identify excited sates of $A$, $B$ series, with the cost of a denser $k$-mesh and longer computational times.

\section{Results \label{sec:III results}}

\subsection{Computational system \label{sec:III:computation}}

To explore the influence of Rashba and TW, we consider the system air/MoTe$_2$/substrate,  where  MoTe$_2$ is described by the band gap $E_\mathrm{g}=$1.4 eV, 
which
matches the experimentally observed peak position at $\sim$1.2 eV for MoTe$_2$~\cite{Arora2016:NL}. The Fermi velocity is $ 
v_\mathrm{F}=
3.28\times10^5$ m/s, and $\lambda_c=-18$ meV and $\lambda_v=110$ meV determine the SOC splittings [recall Eq.~(\ref{eq:H0})], while the lattice constant is $a_0=3.52$ \r{A} for MoTe$_2$~\cite{Qi2015:PRB,Scharf2017:PRL}. As in Eq.~(\ref{eq:HTW}), TW is parameterized by  $\kappa=-1.1$ eV\AA~\cite{Kormanyos2013:PRB,Kormanyos2015:2DM}. 
The presence of a substrate leads to Rashba SOC, which can be as large as $\lambda_\mathrm{R}=72$ meV~\footnote{In Ref.~[\onlinecite{Qi2015:PRB}], $\lambda_R$ is obtained through comparing the band gap and splitting from the low-energy effective Hamiltonian with first-principles calculations.}, which guides the range of Rashba SOC we examine. Specifically,  $\lambda_\mathrm{R}=50$ meV is large enough to clearly show the emergent excitons. Additionally, a substrate influences the background dielectric constant, here taken as $\varepsilon=12.45$ to be consistent with a HfO$_2$ substrate~\cite{Wilk2001:JAP},
which is used with various TMDs~\cite{McDonnell2013:ACSN,Zhang2019:APE} and $r_0=6.3$ nm. Together,  $\varepsilon$ and $r_0$ characterize the polarization of our computational system~\cite{Scharf2017:PRL,Berkelbach2013:PRB,Berkelbach2015:PRB}.

For our numerical calculations, a $N \times N $ $k$-grid with a spacing of $\Delta k=2\pi/(Na_0)$ in each direction is used. We limit our calculation by an upper cutoff energy $E_\mathrm{cut}=E_\mathrm{g}/2+500$ meV.
To ensure convergence in our numerical calculations, we have used $N=250$. A broadening $\Gamma=5$ meV is used for the spectrum, which is large enough for a smooth curve, while small enough to identify possible new excitonic features arising from Rashba SOC.

\subsection{Rashba SOC and emergent excitons\label{sec:III:Rashba}}
\begin{figure}[h]
	\includegraphics[width=0.47\textwidth]{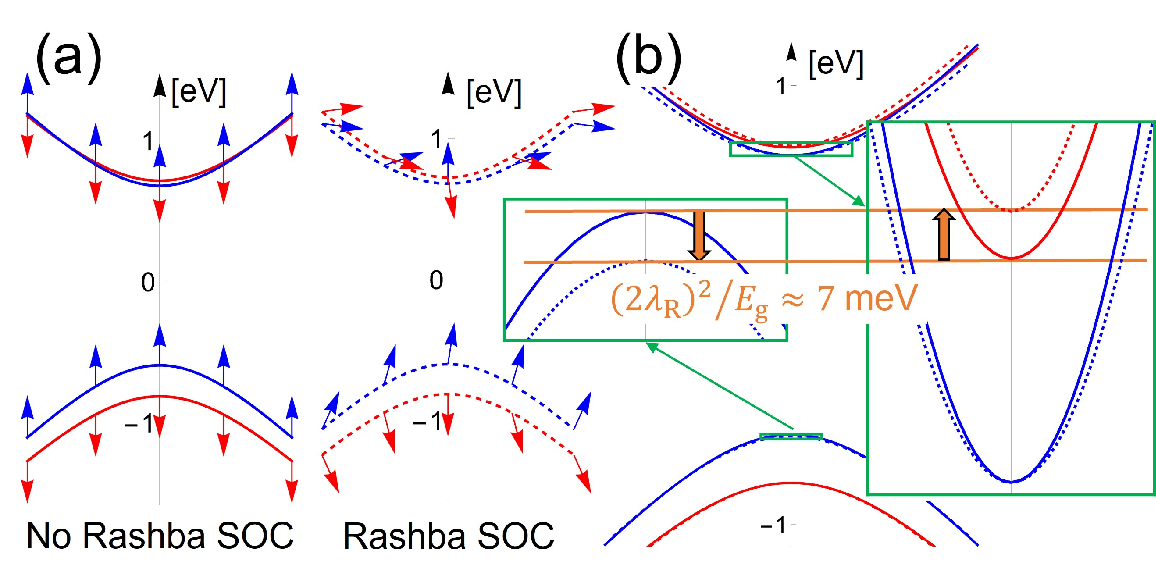}
	\caption{\label{fig:FR2} (a) Schematic single-particle band structure and spin textures (denoted by arrows) in the $K$ valley
with Rashba SOC strength $\lambda_\mathrm{R}=0, 50\;$meV.
	The other parameters are given in Sec.~\ref{sec:III:computation}. (b) Magnified band edges with the indicated Rashba-induced blue shift, where $E_\mathrm{g}$ is the band gap. 
	}
\end{figure}
\subsubsection{Single-particle description}
While the absorption spectra in TMDs are typically dominated by excitons and therefore are not captured by the single-particle description, some of the modifications of excitonic properties with Rashba SOC can already be inferred from the trends present at the single-particle picture. To show this, we compare the band structures 
and the corresponding spin textures at the $K$ valley with and without Rashba SOC, showing a simple situation where the collinear spins ensure that spin is a good quantum number.  In Fig.~\ref{fig:FR2}(a) we see that the interplay between intrinsic and Rashba SOC leads to noncollinear spin textures.
The resulting absence of a single spin quantization axis suggests that the optical selection rules for dipole transitions are relaxed and that initially forbidden dark excitons
can become bright and optically allowed. 

In our case, we have different types of ``dark'' excitons: the spin-forbidden excitons ($A'$ and $B'$) and $p$-shell spin-allowed excitons ($A$ and $B$), which can be active, though, in two-photon processes \cite{Ivchenko:2005}. Relaxing the spin quantization axis, in general, can lead to the optical brightness of both of these excitonic states. For instance, within the single particle picture, the transitions between spin-split energy levels with opposite spin states can be allowed in the presence of finite SOC due to the electric dipole spin resonance \cite{Rashba:SPSS}. This process can make $A'$: and $B'$: $s$-shell excitons optically bright; arising microscopically from SOC-assisted interband spin-flip transitions \cite{Inglot2014:PRB}. However, as we demonstrate below, at sufficiently large magnitudes of SOC, the dominant role for the absorption spectra comes from a different 
effect: the mixing of $s$-shell 
and $p$-shell excitons.

This situation is analogous to TMDs in the absence of Rashba SOC on a magnetic substrate with an in-plane magnetization~\cite{Scharf2017:PRL}.
From the magnified view of the CB and VB near the $K$ valley, in Fig.~\ref{fig:FR2}(b) we can anticipate two more trends for excitons with the increase in $\lambda_\mathrm{R}$:
(i) The absorption peaks will blue-shift, which can be understood as a result of the indicated moving down/up of VB2/CB2, by $(2\lambda_\mathrm{R})^2/E_\mathrm{g}$.  (ii) The heights of the 
excitonic $A$ peak will slightly increase since CB1 and VB2 become flatter. 
The states closer to band edges dominate both single-particle and many-body absorption spectra~\cite{Yu:2010}. Therefore,
a higher density of states around band edges caused by flatter bands leads to an enhanced absorption for the corresponding transition.

\subsubsection{Absorption spectrum: Emergent bright excitons}

\begin{figure}[t]
	\vspace{-0.5cm}
	\includegraphics[width=0.47\textwidth]{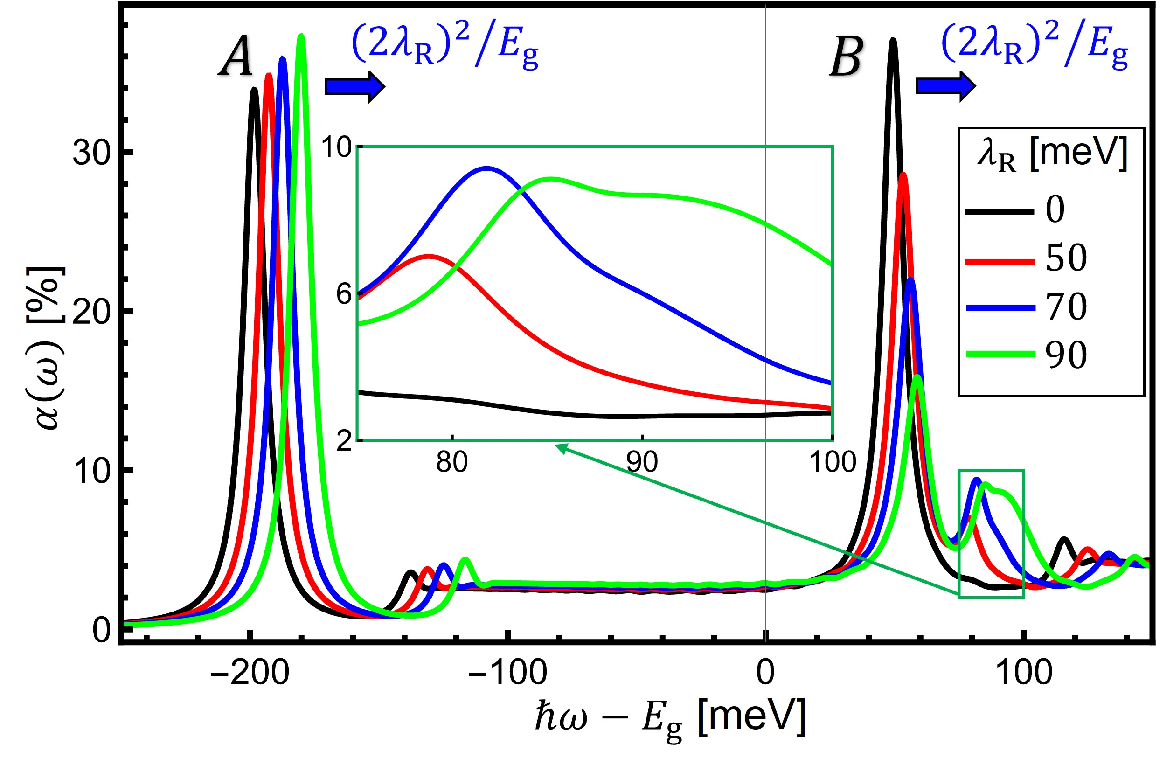}
	\caption{\label{fig:FR3} Absorption spectra for an air/MoTe$_2$/substrate system with different strengths of Rashba SOC. Blue arrows:
		the energy shifts of $A$ and $B$ excitons, expected from the single-particle analysis.
		The inset highlights the emergence of Rashba-induced peaks. The other parameters are given in Sec.~\ref{sec:III:computation}.}
\end{figure}

We next turn to the calculation of absorption spectra from Eq.~(\ref{eq:abs}) using the BSE [Eq.~(\ref{eq:BSE})] to include the effects of strong Coulomb interaction
and explore various trends in the excitonic  peaks with Rashba SOC. The importance of using the BSE can be readily seen from the comparison
with our results for a single-particle absorption, which are calculated for the same air/MoTe$_2$/substrate system, discussed further  in the Appendix (Fig.~\ref{fig:FR8}).
At the single-particle level, excitonic peaks are absent. The exciton binding energies, $\approx 70\;$meV (Fig.~\ref{fig:FR7}) are responsible for the red shift of the $A$ and $B$ peaks, as compared to what would be expected from the corresponding energies of the CB and VB. 
For our considered system, this red shift is much larger than the blue shift due to Rashba SOC,  which has been inferred from the single-particle analysis in Fig.~\ref{fig:FR2}(b) and marked with arrows in Fig.~\ref{fig:FR3}.

Several striking features can be readily seen in Fig.~\ref{fig:FR3}. First of all, we do not observe a noticeable rise of the absorption due to the brightness of the $s$-shell spin forbidden excitons $A'$:$1s$ and $B'$:$1s$ that might have been expected from the single-particle interband spin-flip mechanism \cite{Inglot2014:PRB}. Instead, we see an additional emergent absorption peak with energy slightly higher than the $B$:$1s$ exciton but smaller than the $B$:$2s$ transition. Its evolution is highlighted in the inset, showing that with increasing $\lambda_\mathrm{R}$ its height has increased and its position moves to higher energies. Importantly, there is no similar emergent Rashba-induced excitonic peaks close to the $A$:$1s$ excitonic transition. We address this point in the Appendix (Fig. \ref{fig:FR9}). Moreover, unlike only a small increase in the height of the $A$ peak, consistent with what is expected from the single-particle analysis in Fig.~\ref{fig:FR2}(b), we see that with increasing $\lambda_\mathrm{R}$ the spectral weight of the $B$ peak is strongly reduced and transferred to the Rashba-induced peaks as shown in the inset. This difference between $A$- and $B$-exciton series is important for understanding better the microscopic mechanism that dominates the Rashba-induced modification of the optical response.

\begin{figure}[h]
	\includegraphics[width=0.48\textwidth]{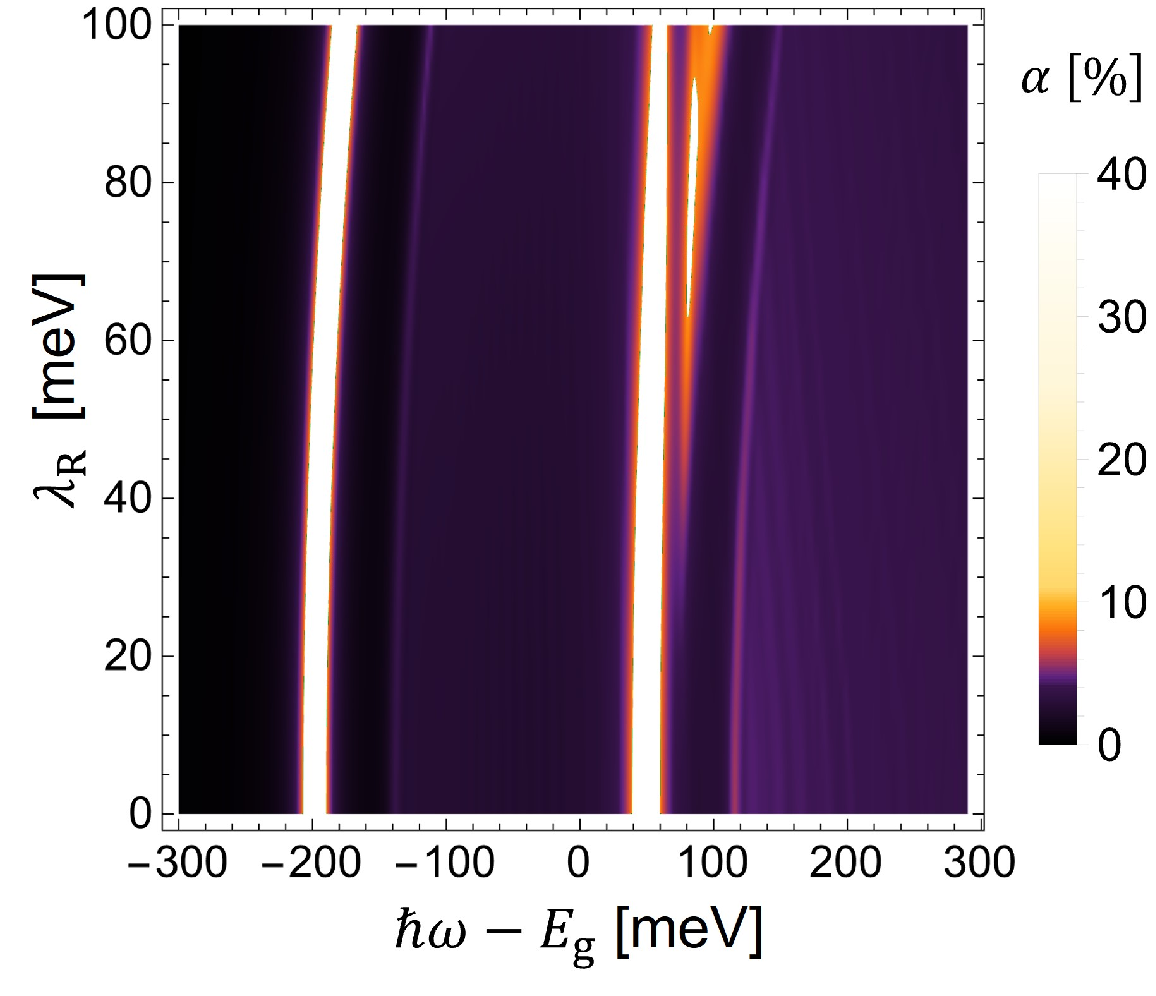}
	\caption{\label{fig:FR4}  Evolution  of the absorption spectrum as Rashba SOC, $\lambda_R$,
		is increased from $0$ meV to $100$ meV. The parameters are the same as in Fig.~\ref{fig:FR3} .}
\end{figure}

For a further view of the evolution of the $A$ and $B$ peaks along with multiple emergent peaks, in Fig.~\ref{fig:FR4} we show a heat map that covers the evolution of the absorption spectrum as  $\lambda_\mathrm{R}$ is increased from $0$ to $100\;$meV. While in Fig.~\ref{fig:FR3} we can already see that the $B$ peak is accompanied by several satellite peaks, 
in Fig.~\ref{fig:FR4} from a gradual decrease in the amplitude of the $B$ peak it can be  seen that it dissolves into several peaks. As $\lambda_\mathrm{R}$ increases,
these emergent satellite peaks, at energies above the $B$ peak, have a nonmonotonic spectral weight and position. This behavior leads to the overall broadening, 
as compared to the original $B$ peak at $\lambda_\mathrm{R}=0$. At large $\lambda_\mathrm{R}$ there is a slightly decreased separation between $A$ and $B$ peaks, unlike the equal blue shifts expected from the single-particle analysis and denoted in Fig.~\ref{fig:FR3}.

\subsubsection{Origin of emergent excitons: Mixing of excitonic states}

While our previous analysis of the absorption spectra has focused on the $k$-space study of the excitonic features, complementary information is obtained from the real-space
descriptions and the possibility to classify their symmetry, as suggested in Sec.~\ref{sec:II:BSE}. 
Combining $k$- and real-space descriptions of the excitons in CrX$_3$ has recently provided valuable information on the role of hybridization
in the excitonic properties by changing X from Cl to Br and I~\cite{Acharya2022:npj2D}.
Furthermore, the real-space exciton envelope function Eq.~(\ref{eq:wave_total}) can be studied in detail by decomposing it into four parts according to the combination of subband indices $c$ and $v$ as
\begin{multline}
	\Psi^{S, \tau}(\bm{r_e}-\bm{r_h})= \Psi_A^{S, \tau}(\bm{r_e}-\bm{r_h})+\Psi_{A'}^{S,\tau}(\bm{r_e}-\bm{r_h}) \\
	+\Psi_B^{S, \tau}(\bm{r_e}-\bm{r_h})+\Psi_{B'}^{S, \tau}(\bm{r_e}-\bm{r_h}),
	\label{eq:WavePartialSum}
\end{multline}
where $A$,  ($A'$, $B$ or $B'$) corresponds to the summation over terms with $v, c=$ VB2 (VB2, VB1 or VB1), CB1 (CB2, CB2 or CB1) in Eq.~(\ref{eq:wave_total}).  For example, $\Psi_{A}^{S, \tau}(\bm{r_e}-\bm{r_h})=
\sum_{\bm{k}} {\mathcal A}_{vc\bm{k}}^{S\tau} \psi^\tau_{c\bm{k}}(\bm{r_e})\psi_{v\bm{k}}^{\tau *}(\bm{r_h})$ with $v=\mathrm{VB2}, c=\mathrm{CB1}$.

\begin{figure}[t]
	\includegraphics[width=0.48\textwidth]{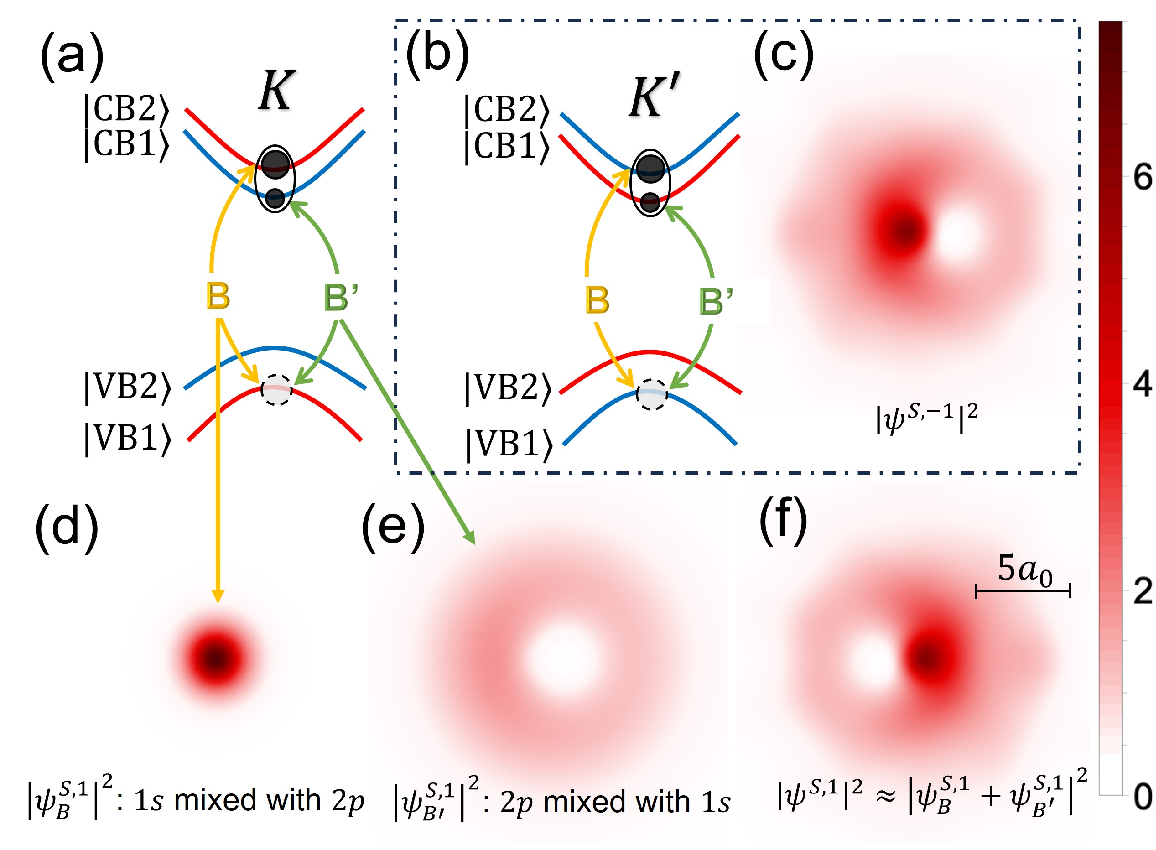}
	\caption{\label{fig:FR5} 
		(a),(b) Schematic illustration of the mixing of $B$  and $B'$ excitonic states, formed from both CBs and VB1, in the $K$ and $K'$ valley. 
		(c),(f)  Modulus squared of the real-space wave function as a function of $\bm{r_e}-\bm{r_h}$ for the Rashba-induced exciton in the $K'$ and $K$ valley with the energy $\Omega^{S,-1}=\Omega^{S,1}=1.4789 \;$eV, from Eqs.~(\ref{eq:BSE}) and (\ref{eq:wave_total}).
		(d),(e)  $|\psi_B^{S,1}|^2$( $|\psi_{B'}^{S,-1}|^2$) as defined in Eq.~(\ref{eq:WavePartialSum}):
		Modulus squared of $B$($B'$) excitonic states formed from CB2 (CB1) and VB1 coupling, dominated by $1s$ ($2p$) state, in the $K$ valley.
	}
\end{figure}

In the absence of Rashba SOC, each exciton is formed from only one CB and VB and can be labeled by the spin configuration of those bands. This is clear from the ground-state ($1s$)  $A$ or $B$ excitons [recall Fig.~\ref{fig:FR1}(b)], which are formed from the electron-hole coupling exclusively from CB1 and VB2 or CB2 and VB1. They satisfy $\Psi^{S, \tau}(\bm{r})= \Psi_A^{S, \tau}(\bm{r})$ and $\Psi^{S, \tau}(\bm{r})= \Psi_B^{S, \tau}(\bm{r})$ separately with the exciton envelope 
wave function having either $s$-shell or $p$-shell symmetry, the latter is characterized by nonzero angular momentum.
For nonzero $\lambda_\mathrm{R}$, we focus on the emerging transition with the energy corresponding to the new peak in Fig.~\ref{fig:FR3}, which we call a Rashba-induced exciton. In Fig.~\ref{fig:FR5} we present the modulus squared of the real space envelope wave function of the Rashba-induced exciton. We see that for $\lambda_\mathrm{R}=50\;$meV, $|\Psi^{S, \tau}|^2$ is neither a perfect $s$- nor a $p$- shell wave. It is asymmetrical along the $x$ direction in both valleys. It is interesting to note that $|\Psi^{S, 1}|^2$ and $|\Psi^{S, -1}|^2$ are related by reflection with respect to the $y$ axis, a property that can be easily verified through step-by-step derivation starting from the effective Hamiltonian 
in Eq.~(\ref{eq:Htot}). 

The exciton envelope function $|\Psi^{S, \tau}|^2$ of a Rashba-induced exciton, which deviates from a conventional non-Rashba $A$/$B$ exciton, can be understood by examining its four constituents, that is $\Psi^{S, \tau}_A$,$\Psi^{S, \tau}_{A'}$,$\Psi^{S, \tau}_B$ and $\Psi^{S, \tau}_{B'}$. By comparing their contributions, we find that $B$ and $B'$ excitonic states dominate, while a negligible amount of continuum $A$ and $A'$ states also contribute. Thus, a Rashba-induced exciton can be understood as the result of $B$ and $B'$ excitonic states mixing as summarized schematically in 
Figs.~\ref{fig:FR5}(a) and \ref{fig:FR5}(b).
Furthermore, $B$ and $B'$ excitonic states are $1s$- and $2p$-like, respectively, as shown by plots of their modulus squared 
wave functions
in Figs.~\ref{fig:FR5}(d) and  \ref{fig:FR5}(e).
We can clearly see in Fig.~\ref{fig:FR5}(e) that the $B'$ excitonic state is not perfectly $2p$-like but asymmetric due to $1s$ admixture.
Similarly, the $B$ excitonic state is predominately $1s$ with a small $2p$ contribution.

The Rashba-induced excitonic state-mixing discussed above can be understood by careful examination of the BSE. Without Rashba, excitons consist only of excitations from one valence band to one conduction band because the kernel, Eq.~(\ref{eq:KerneltEffectiveModel}), of the BSE, Eq.~(\ref{eq:BSE}), is diagonal, i.e. proportional to $\delta_{cc'}$ $\delta_{v,v'}$', where $\delta_{ij}$ denotes Kronecker symbols. As Rashba SOC is turned on, the kernel is no longer diagonal, and excitations between various valence and conduction bands couple to each other. This then leads to a mixing of previously decoupled states, which are close in energy, for example 
$B$:$1s$ (originally only from the transitions VB1 to CB2 for $\lambda_\mathrm{R}=0$) and $B'$:$2p$ excitons (originally only from the transitions VB1 to CB1 for $\lambda_\mathrm{R}=0$). The mixing is the strongest for the states having close energies. For this reason, in particular, the Rashba-induced excitonic state is not visible for the $A$-series of excitons. The
$A$:$1s$ 
 and 
$A'$:$2p$
states have larger energy mismatch. Likewise, the individual contributions $B$ (CB2-VB1) and $B'$ (CB1-BV1) also acquire small contributions from each other, while their main character is still mainly $s$- and $p$-like, respectively. 

The obtained Rashba-induced exciton represents an additional microscopic mechanism contributing to the important class of excitonic $s$-shell and $p$-shell mixing phenomena. Let us emphasize 
several key physical ingredients, differentiating it from the previously studied scenario due to the symmetry reduction, considered in 
Ref.~\cite{Glazov2017:PRB}.
That prior work focused on the spin-allowed excitonic states (the dark excitons due to opposite spin states were neglected) and demonstrated that the $s$-$p$ mixing in the same valley can be achieved when accounting for the admixture of the remote bands. Using the $\bm{k}\cdot\bm{p}$ method beyond the axially symmetric model, it was shown that it is the admixture parameter (related to the interband matrix element of the momentum operator) with remote bands that gives rise to the mixing of $2s$ and $2p$ states and makes $p$-excitons optically active. In our case, the symmetry reduction is realized by means of the proximity-induced modification of the SOC, resulting in the appearance of the Rashba term. In our scheme, the $s$-$p$ mixing can be obtained without accounting for the remote bands and using only the minimal effective model of Dirac bands in TMDs.  Interestingly, the Rashba-induced exciton is combined from $1s$ and $2p$ states with spin-allowed and spin-forbidden transitions, which contrasts the remote-band mechanism of the $s$-$p$ mixing from Ref.~\cite{Glazov2017:PRB}, requiring the same spin states.

Our analysis also confirms that the emergent excitons can be understood as the brightening
of the dark states.  However, compared to the previously studied case with a magnetic substrate where only the $1s$ excitons are brightened with an in-plane magnetization~\cite{Scharf2017:PRL}, there are several differences. With Rashba SOC-induced spin textures, a $B'$:$2p$ (CB1-VB1) exciton remains optically forbidden, while it is the mixture with a $B$:$1s$ (CB2-VB1) optically allowed state that brightens this exciton. 
Interestingly, even without Rashba SOC, mixing of  $s$-$p$ states and an unusual excitonic structure was attributed to the presence of orbital textures in semiconductor nanostructures~\cite{Lee2014:PRB}.

\begin{figure*}[t]
	\includegraphics[width=0.8\textwidth]{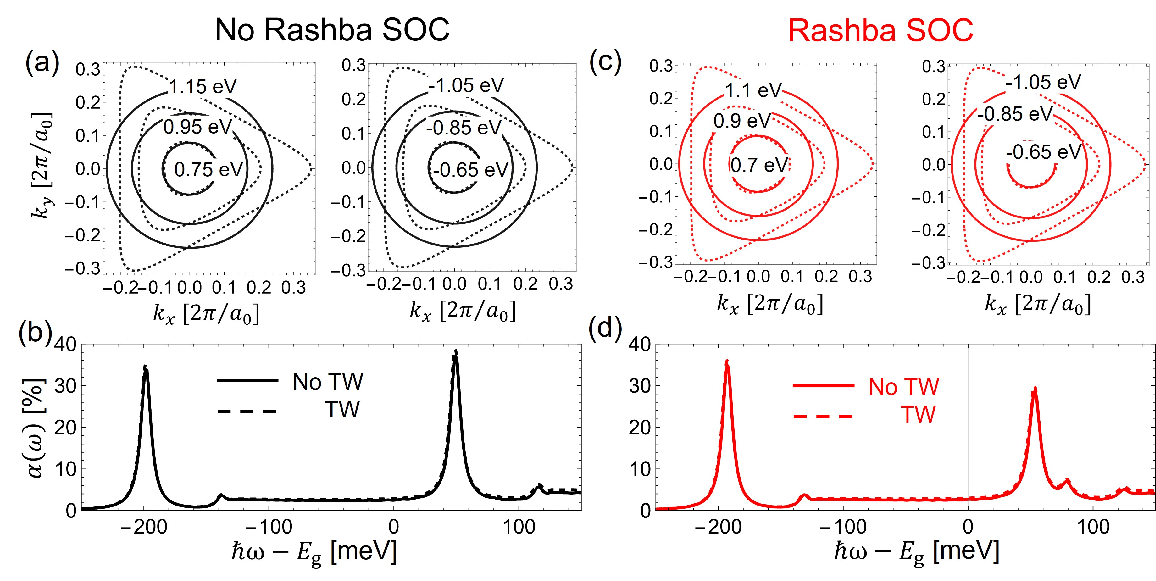}
	\caption{Effect of trigonal warping (TW) on the band structures in the $K$ valley and on the absorption spectra, in the absence of Rashba SOC (a) and (b).
		A similar analysis in the presence of Rashba SOC, $\lambda_R=50\;$ meV in (c) and (d). Solid (dashed) lines represent the absence (presence) of TW.
		In (a), without Rashba SOC,  the energy contours of CB1 and VB2 are shown separately and labeled by the corresponding positive and negative energies.
		Similarly, in (c), they are shown in the presence of Rashba SOC.}  
	\label{fig:FR6} 
\end{figure*}
\subsection{Trigonal warping}

Similar to exploring the role of Rashba SOC on the absorption spectra, we also consider wheather the presence of TW leads to observable changes.
We again use BSE calculations where TW is described by Eq.~(\ref{eq:HTW}). Even without considering a substrate, TW is inherent to the band structure in many 2D materials, including graphene~\cite{CastroNeto2009:RMP}, boron nitride~\cite{Galvani2016:PRB} and TMD MLs~\cite{Kormanyos2015:2DM}. The presence of TW leads to a threefold 
anisotropy in the band structure that directly influences different optical properties. For example, strong optical nonlinearities and second harmonic generation are associated with TW as it breaks
rotational symmetry~\cite{Saynatjoki2017:NC}. Furthermore, TW can modify the chiral optical selection rules by changing the single-particle velocity matrix 
$\mathcal{D}_{vc\bm{k}}^{\tau,\pm}$, and cause mixing of excitonic states~\cite{Gong2017:PRB,Glazov2017:PRB}. 
 
To examine the influence of TW, we first show in Figs.~\ref{fig:FR6}(a) and \ref{fig:FR6}(c) how it changes CB and VB isoenergy contours without and with Rashba SOC.
While both in CB and VB there is a striking change from the isotropic band structure to a threefold anisotropy with TW, a further inclusion of Rashba SOC leads
only to modest changes. Turning to the absorption spectra, our calculations in Figs.~\ref{fig:FR6}(b) and \ref{fig:FR6}(d) show that for realistic 
TW strengths there are negligible changes on the excitonic spectra. In the absence of TW, spin-valley locking
implies that at $k=0$ each valley is associated with interband optical transitions with only a single helicity of light, while this locking is reduced at $k\neq0$,
allowing admixture with the other helicity~\cite{Xiao2012:PRL}. With TW, spin-valley locking is further reduced.

For the considered calculations, where we include the contributions of the $K$ and $K'$ valleys, we conclude that neglecting the TW effects is an accurate approximation for the absorption spectra in ML TMDs. This conclusion is not a surprise considering that low-energy exciton 
wave functions in ML TMDs are highly localized around the $K/K'$ points~\cite{Qiu2016:PRB}. In contrast, TW is more pronounced far away from the $K/K'$ points. This agrees
with our conclusions that the influence of TW in the study of excitons is negligible,
while it can be noticeable in the single-particle band structure. 

\subsection{Binding energy}
\begin{figure}[h]
	\includegraphics[width=0.47\textwidth]{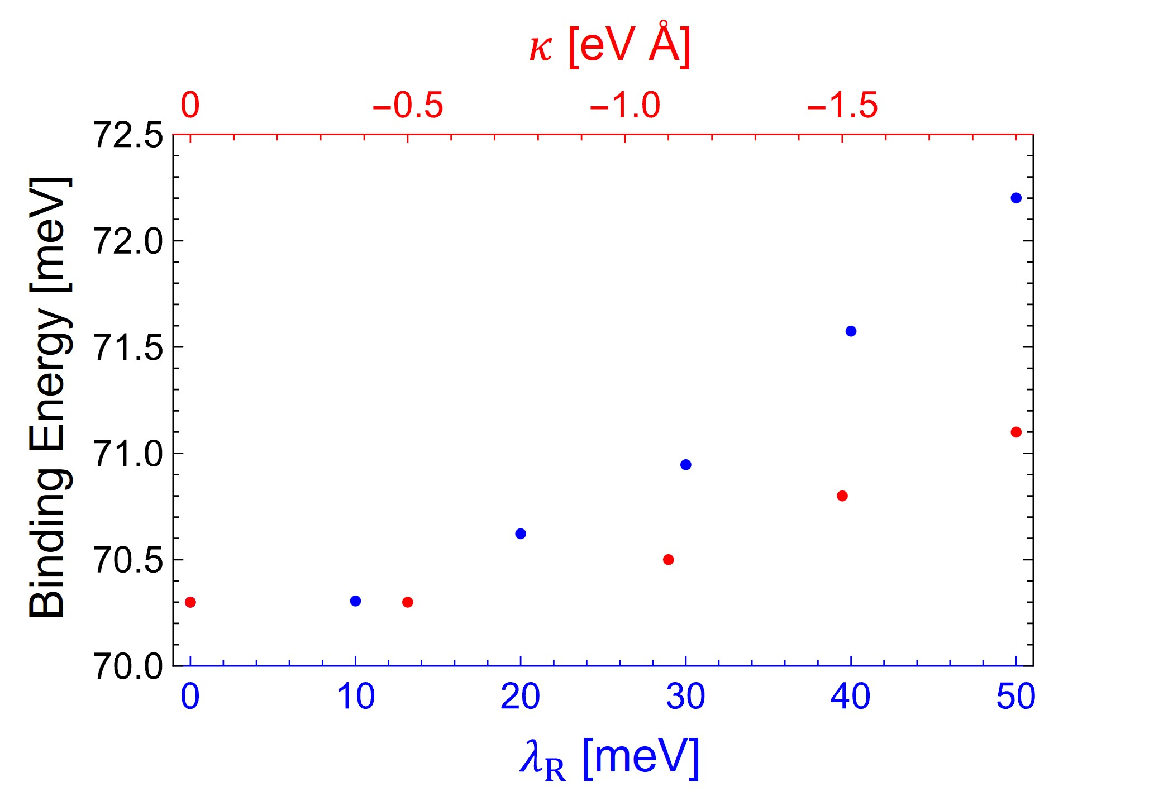}
	\caption{\label{fig:FR7} Evolution of the binding energy for the ground state ($1s$) $A$ exciton with TW and Rashba SOC,
	from Eq.~(\ref{eq:bind}).
	}
\end{figure}

One of the hallmarks of ML TMDs is the occurrence of tightly bound excitons, characterized by binding energies
that can exceed $100$ meV~\cite{Wang2018:RMP}. To investigate the effects of Rashba and TW on the stability of excitons, we explore the change of the ground state ($1s$) $A$ exciton binding energy in Fig.~\ref{fig:FR7}. The binding energy of an exciton $S$ in the $\tau$ valley is calculated as the difference in energy between a free and bound electron-hole pair,
\begin{equation}
E_\mathrm{b}=\epsilon_c^\tau-\epsilon_v^\tau-\Omega^{S,\tau},
\label{eq:bind}
\end{equation}
where, as defined in Eq.~(\ref{eq:BSE}),  $\epsilon_c^\tau$ ($\epsilon_v^\tau$) is the conduction (valence) band edge, and $\Omega^{S,\tau}$ is the exciton energy in the state $S$ in the $\tau$ valley.
For realistic values of Rashba SOC and TW, the binding energy, $E_\mathrm{b}$, is increased by up to about $2\;$meV from the value of $70.3\;$meV, which was obtained without Rashba SOC and TW. This resulting small effect of Rashba SOC and TW on $E_\mathrm{b}$ suggests that the single-particle description of the changes in
the band-edge position can accurately capture the shift of the excitonic peaks. 

\section{Conclusions and Outlook\label{sec:IV Conclusions}}
We have used a low-energy effective Hamiltonian and the resulting four-band model to study the influence of Rashba SOC and TW on the 
optical properties of ML TMDs. Both Rashba SOC and TW were found to strongly modify the single-particle properties through creating noncollinear spin textures and forming a threefold 
anisotropy in the isoenergy contours, respectively. In contrast, by using the Bethe-Salpeter equation to calculate the optical absorption, we found a significant change for 
Rashba SOC, which leads to emergent excitons by coupling different conduction bands and mixing of $1s$ and $2p$ excitonic states. With TW the calculated changes on the absorption spectra are negligible.

The relative importance of Rashba SOC, as compared to TW in the study of excitons comes from 
the localization of excitons around band edges $K$/$K'$~\cite{Qiu2016:PRB}. 
The corresponding exciton 
wave functions are mainly affected by the dispersion around the $K$/$K'$ points. The dispersion around the $K$/$K'$ points, in turn, is strongly affected by Rashba SOC, which breaks the 
collinear spin alignment and introduces energy shifts at the band edges (recall Fig.~\ref{fig:FR2}). TW is, on the other hand, more pronounced far away from the $K$/$K'$ points, and thus negligible in the study of excitons.  
 
However, this situation, where the neglect of TW provides an accurate description of optical properties is not universal. For example, 
TW can lead to strong nonlinearities~\cite{Mahmoudi2022:PRB,Saynatjoki2017:NC,Zhumagulov2022:PRB} and important changes in the optical
response when only a single valley is considered, or if there is an in-plane applied electric field. By changing such an electric field, which alters the overlap between conduction and
valence bands~\cite{Wang2020:PRR}, one can control the degree of circularly polarized light, which has been experimentally demonstrated in ML WSe$_2$,~\cite{Zhang2014:S,GonzalezMarin2022:NC}. 
This mechanism could provide a modulation of spin polarization, suitable for proposed TMD-based
 spin-lasers~\cite{Lee2014:APL}. 

Given the importance of excitons in the optical properties of ML TMDs, we expect
that our analysis for the influence of Rashba SOC and TW will stimulate further studies as they are inherent to many systems. Specifically, the predicted presence
of emergent excitons could be experimentally tested in Janus TMDs, as they support sufficiently strong Rashba SOC, for example, $\alpha_\mathrm{R}$ in 
MoSSe (WSSe) is 77$\;$meV (158$\;$meV)~\cite{Tang2022:PSSB}. 
It would be interesting to explore the role of strain~\cite{Zollner2019:PRB} in the presence of Rashba
SOC,
as their interplay would change the optical response.
Our work can also be viewed as a guidance to the analysis of various proximity effects, such as those due to magnetic or charge density waves substrates~\cite{Joshi2022:APLM,Scharf2017:PRL,Chirolli2020:PRB}, 
which could directly modify the underlying excitonic features and be influenced by Rashba SOC and TW.

\section{Acknowledgments}
This work is supported by U.S. Department of Energy, Office of Science, Basic Energy Sciences under Award No. DE-SC0004890, National Natural Science Foundation of China under award No. 12104118 and
the UB Center for Computational Research.

\appendix*
\section{Single-particle absorption and the position of Rashba-induced peaks}

\begin{figure}[h]
	\includegraphics[width=0.48\textwidth]{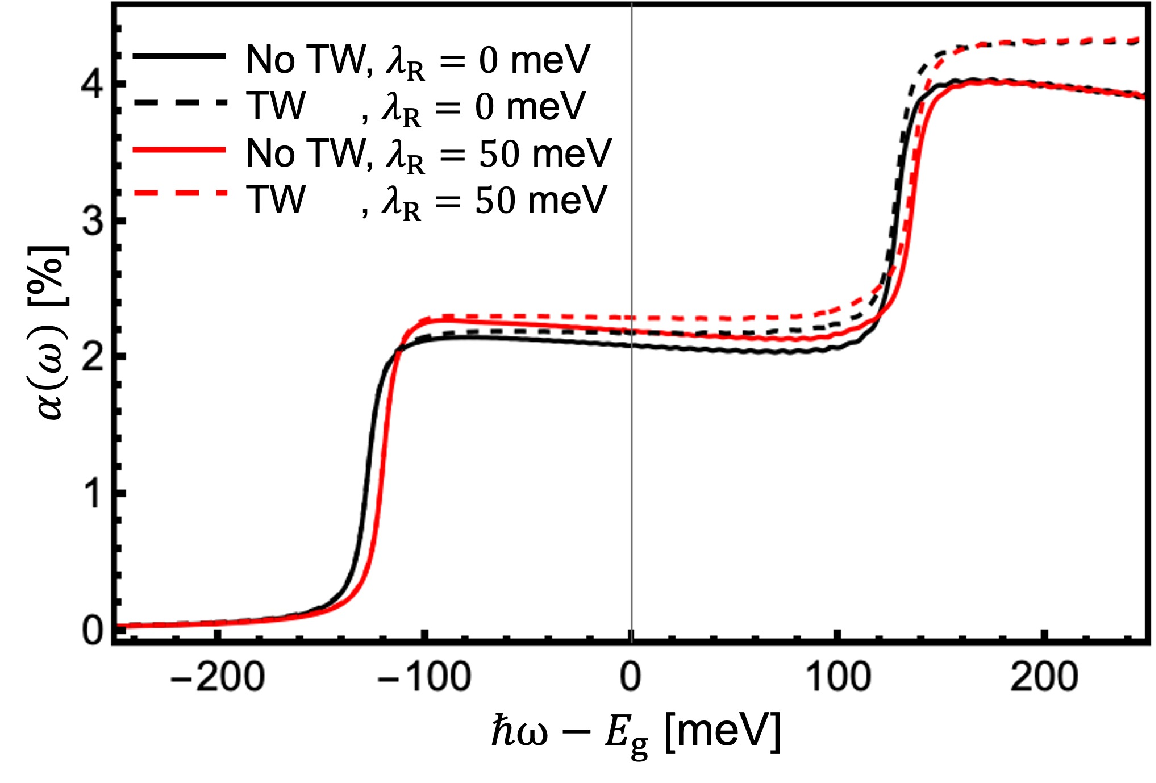}
	\caption{\label{fig:FR8} Single-particle absorption spectra without and with Rashba SOC and TW. The remaining parameters  correspond to those from Fig.~\ref{fig:FR3}.
	}
\end{figure}

By excluding Coulomb interaction, the resulting absorption spectra for air/MoTe$_2$/substrate in Fig.~\ref{fig:FR8}
show pronounced differences compared to previous many-body
results obtained from BSE in Fig.~\ref{fig:FR3}.  Instead of excitonic $A$ and $B$ peaks, there are two jumps in the absorption, corresponding to the transitions between
CB1 and VB2 and CB2 and VB1, respectively. These jumps are at different positions from the $A$ and $B$ peaks as they are not red-shifted by the binding energies, 
$E_b$, as shown in Fig.~\ref{fig:FR7}, while the overall maximum in the absorption is nearly an order of magnitude smaller than in Fig.~\ref{fig:FR3}.
The presence of Rashba SOC and TW leads to only small modifications in the single-particle absorption.

\begin{figure}[t]
	\includegraphics[width=0.47\textwidth]{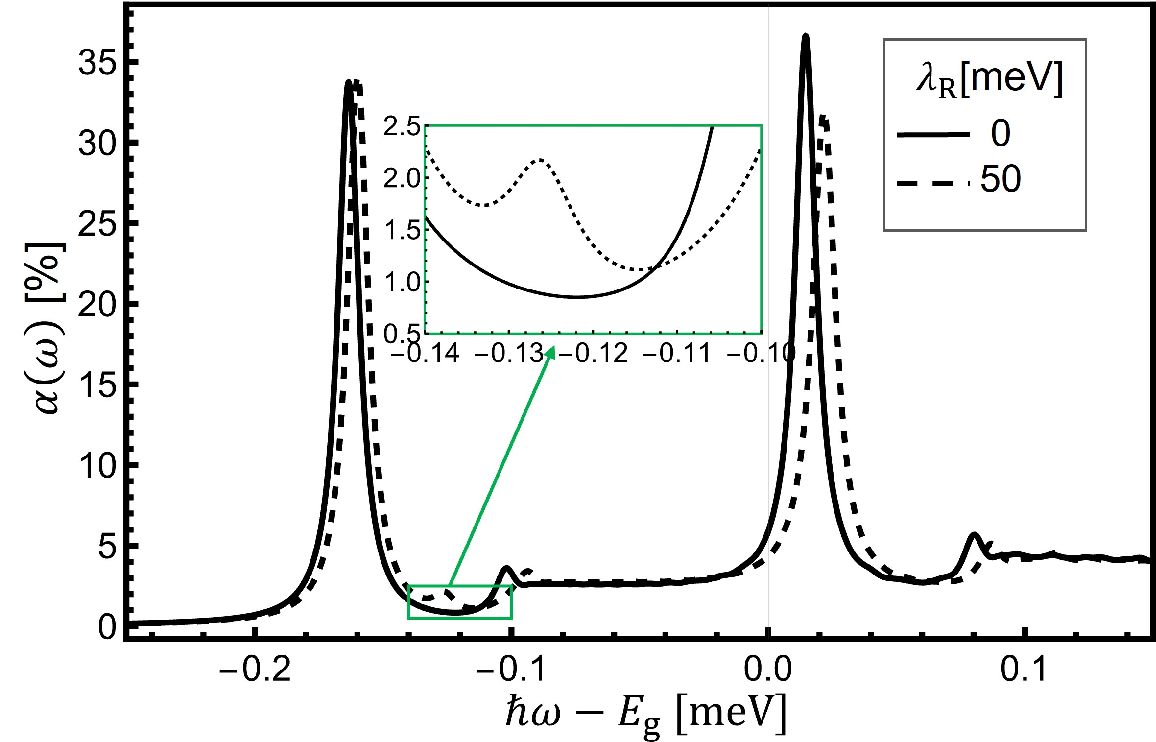}
	\caption{\label{fig:FR9} Absorption spectrum without and with Rashba SOC ($\lambda_R=0, \; 50\;$meV).
	 In each case
	 the intrinsic CB SOC, $\lambda_c=18\;$meV is reversed compared to the previous calculations. The inset highlights an emergent
	 excitonic peak near the A exciton. The remaining parameters are the
	 same parameters as in Fig.~\ref{fig:FR3}. 
	}
\end{figure}

One striking feature of the emergent extra excitonic peaks is that they appear only after the $B$ peak, but not the $A$ peak, as can be seen in Figs.\ref{fig:FR3} and \ref{fig:FR4}.
A simple understanding of this property is that an energy match is necessary for the admixture of various states which form these emergent excitons. 
For example, the ground state of the $B$ exciton has a larger energy than that of the $B'$ exciton. Hence it is possible to have an admixture between  $1s$ of the $B$ exciton and $2p$ of the $B'$ exciton. In contrast, due to the band ordering, determined by the intrinsic SOC parameters $\lambda_c$ and $\lambda_v$, there is no such 
an admixture between  $1s$ of the $A$ exciton and $2p$ of the $A'$ exciton. Our understanding is verified in Fig.~\ref{fig:FR9}, where
our previous calculations are repeated by artificially reversing the band ordering of the CBs by 
imposing $\lambda_c\rightarrow -\lambda_c$ which leads to an extra peak after the $A$ peak rather than the $B$ peak.  

\bibliography{TWR}

\end{document}